\documentclass[twoside,twocolumn,9pt]{article}
\usepackage{extsizes}
\usepackage[super,sort&compress,comma]{natbib} 
\usepackage[version=3]{mhchem}
\usepackage[left=1.5cm, right=1.5cm, top=1.785cm, bottom=2.0cm]{geometry}
\usepackage{balance}
\usepackage{mathptmx}
\usepackage{sectsty}
\usepackage{graphicx} 
\usepackage{lastpage}
\usepackage[format=plain,justification=justified,singlelinecheck=false,font={stretch=1.125,small,sf},labelfont=bf,labelsep=space]{caption}
\usepackage{float}
\usepackage{fancyhdr}
\usepackage{fnpos}
\usepackage[english]{babel}
\addto{\captionsenglish}{%
  
}
\usepackage{array}
\usepackage{droidsans}
\usepackage{charter}
\usepackage[T1]{fontenc}
\usepackage[usenames,dvipsnames]{xcolor}
\usepackage{setspace}
\usepackage[compact]{titlesec}
\usepackage{hyperref}
\usepackage{soul}

\usepackage{epstopdf}

\definecolor{cream}{RGB}{222,217,201}

\begin{document}

\pagestyle{fancy}
\thispagestyle{plain}
\fancypagestyle{plain}{
\renewcommand{\headrulewidth}{0pt}
}

\makeFNbottom
\makeatletter
\renewcommand\LARGE{\@setfontsize\LARGE{15pt}{17}}
\renewcommand\Large{\@setfontsize\Large{12pt}{14}}
\renewcommand\large{\@setfontsize\large{10pt}{12}}
\renewcommand\footnotesize{\@setfontsize\footnotesize{7pt}{10}}
\makeatother

\renewcommand{\thefootnote}{\fnsymbol{footnote}}
\renewcommand\footnoterule{\vspace*{1pt}%
\color{cream}\hrule width 3.5in height 0.4pt \color{black}\vspace*{5pt}} 
\setcounter{secnumdepth}{5}

\makeatletter 
\renewcommand\@biblabel[1]{#1}            
\renewcommand\@makefntext[1]%
{\noindent\makebox[0pt][r]{\@thefnmark\,}#1}
\makeatother 
\renewcommand{\figurename}{\small{Fig.}~}
\sectionfont{\sffamily\Large}
\subsectionfont{\normalsize}
\subsubsectionfont{\bf}
\setstretch{1.125} 
\setlength{\skip\footins}{0.8cm}
\setlength{\footnotesep}{0.25cm}
\setlength{\jot}{10pt}
\titlespacing*{\section}{0pt}{4pt}{4pt}
\titlespacing*{\subsection}{0pt}{15pt}{1pt}

\fancyfoot{}
\fancyfoot[LO,RE]{\vspace{-7.1pt}\includegraphics[height=9pt]{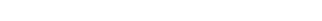}}
\fancyfoot[CO]{\vspace{-7.1pt}\hspace{13.2cm}\includegraphics{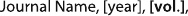}}
\fancyfoot[CE]{\vspace{-7.2pt}\hspace{-14.2cm}\includegraphics{head_foot/RF}}
\fancyfoot[RO]{\footnotesize{\sffamily{1--\pageref{LastPage} ~\textbar  \hspace{2pt}\thepage}}}
\fancyfoot[LE]{\footnotesize{\sffamily{\thepage~\textbar\hspace{3.45cm} 1--\pageref{LastPage}}}}
\fancyhead{}
\renewcommand{\headrulewidth}{0pt} 
\renewcommand{\footrulewidth}{0pt}
\setlength{\arrayrulewidth}{1pt}
\setlength{\columnsep}{6.5mm}
\setlength\bibsep{1pt}

\makeatletter 
\newlength{\figrulesep} 
\setlength{\figrulesep}{0.5\textfloatsep} 

\newcommand{\topfigrule}{\vspace*{-1pt}%
\noindent{\color{cream}\rule[-\figrulesep]{\columnwidth}{1.5pt}} }

\newcommand{\botfigrule}{\vspace*{-2pt}%
\noindent{\color{cream}\rule[\figrulesep]{\columnwidth}{1.5pt}} }

\newcommand{\dblfigrule}{\vspace*{-1pt}%
\noindent{\color{cream}\rule[-\figrulesep]{\textwidth}{1.5pt}} }

\makeatother

\twocolumn[
  \begin{@twocolumnfalse}
{\includegraphics[height=30pt]{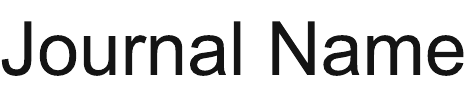}\hfill\raisebox{0pt}[0pt][0pt]{\includegraphics[height=55pt]{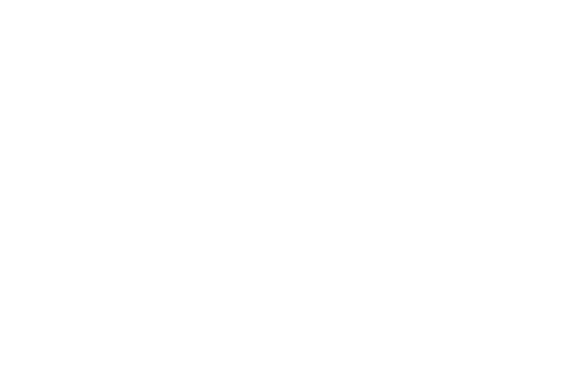}}\\[1ex]
\includegraphics[width=18.5cm]{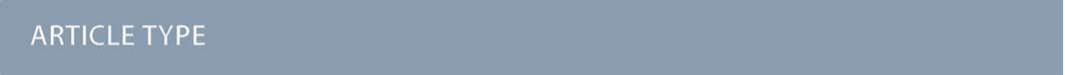}}\par
\vspace{1em}
\sffamily
\begin{tabular}{m{4.5cm} p{13.5cm} }

\includegraphics{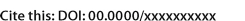} & \noindent\LARGE{\textbf{Nematic liquid crystal flow driven by time-varying active surface anchoring$^\dag$}} \\
\vspace{0.3cm} & \vspace{0.3cm} \\

 & \noindent\large{Seyed Reza Seyednejad$^{\ast}$\textit{$^{a}$} and Miha Ravnik\textit{$^{a,b}$}} \\

\includegraphics{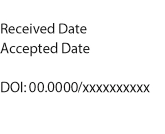} & \noindent\normalsize{
	We demonstrate the generation of diverse material flow regimes in nematic liquid cells as driven by time-variable active surface anchoring, including no-net flow, oscillatory flow, steady flow, and pulsating flow.
	Specifically, we numerically simulate a passive nematic fluid inside a cell bounded with two flat solid boundaries at which the time-dependent anchoring is applied with the dynamically variable surface anchoring easy axis.
	We show that different flow regimes emerge as the result of different anchoring driving directions (i.e. co-rotating or counter-rotating) and relative phase of anchoring driving. The flow magnitude is tunable by cell thickness and anchoring driving frequency. More generally, this work aims towards possible applications of responsive time-variable surfaces, including photonics or synthetic active matter. 
	}

\end{tabular}

 \end{@twocolumnfalse} \vspace{0.6cm}

  ]

\renewcommand*\rmdefault{bch}\normalfont\upshape
\rmfamily
\section*{}
\vspace{-1cm}


\footnotetext{\textit{$^{a}$~Faculty of Mathematics and Physics, University of Ljubljana, Jadranska 19, 1000 Ljubljana, Slovenia; E-mail: seyed-reza.seyednejad@fmf.uni-lj.si}}
\footnotetext{\textit{$^{b}$~Department of Condensed Matter Physics, Jo\v{z}ef Stefan Institute, Jamova 39, 1000 Ljubljana, Slovenia.}}

\footnotetext{\dag~Electronic Supplementary Information (ESI) available: [details of any supplementary information available should be included here]. See DOI: 00.0000/00000000.}



\section{Introduction}

Microfluidics in nematic liquid crystals (NLCs) represents an intersection of soft matter physics and engineering, where the -backflow determines the material response- coupling between the liquid crystalline orientational order and the material flow\cite{brochard1973backflow,svenvsek2001backflow}.
This field explores challenges including how orientation and flow in the nematic liquid crystals be coupled, and how they can be controlled and manipulated using microfabricated channels\cite{giomi2017cross} and surfaces with specific anchoring properties\cite{copar2021,ohzono2012zigzag,luo2018tunable}.
The anisotropic nature of nematic liquid crystals allows for the creation of topologically interesting and functional microstructures that can be dynamically tuned by external stimuli such as electric fields\cite{na2010electrically,fedorowicz2024electrically,kos2020field,emervsivc2019sculpting}, temperature gradient\cite{PhysRevLett.110.048303}, or flow gradients \cite{vcopar2020microfluidic,pieranski1974instability}. These capabilities open up diverse potential applications, ranging from targeted delivery systems\cite{sengupta2013topological} to novel metamaterials\cite{sentker2019self,zhao2007electrically,aplinc2019designed}.

In addition to known mechanisms for pushing fluids out of equilibrium, such as electrophoresis\cite{electrophoresis}, thermophoresis\cite{miralles2013review,PhysicsofFluids2024}, electro-osmosis\cite{wang2009electroosmotic}, electrowetting\cite{choi2012digital}, optical fields\cite{fan2011optofluidic}, and acoustic waves\cite{yeo2014surface},
the dynamics and flow in nematic environments can be driven by internal activity, caused by chemical ingredients\cite{PhysRevE.102.020601,PhysRevLett.123.178003} and biological agents\cite{LivingliquidcrystalsPNAS,zhang2021autonomous}.
The inherently out-of-equilibrium liquid crystal systems are active nematics\cite{science.aah6936,PhysRevLett.122.048004}, which show peculiar topological and elastic properties of such tunable fluids\cite{guillamat2017taming,doostmohammadi2017onset,tan2019topological,kumar2018tunable,kralj2023defect,colen2021machine,bonn2024elasticity}, where the confinement geometry\cite{PhysRevFluids.2.104201} and the channel size\cite{head2024spontaneous} are of basic importance.

In intrinsically passive NLCs, the flow can be a result of the constant entry of elastic distortions generated by a rotating external field\cite{kos2020field}. Elastic distortions can also be brought into the NLC bulk by remotely rotating a solid colloid inside the NLC and making the colloid swim\cite{yao2022topological}, or by oscillating hydrostatic pressure, making spherical bubbles pulsate and swim in the bulk\cite{kim2024symmetrically}.
Moreover, flow in the micro-channel networks can be modulated and guided by using surface anchoring profiles, that are controlled by optical techniques\cite{sengupta2013tuning}.

Our aim here is to drive flow in a passive NLC channel by generating elastic distortions through dynamic surface anchoring. NLC molecular alignments on the interfaces regularly are categorized as planar (parallel), homeotropic (perpendicular), or conical (tilted), which can be experimentally achieved by various physical and chemical techniques\cite{bryan1999weak,uchida1989director}. Polymeric thin film surfaces and monolayer coverage of surfactants are used to guarantee the homeotropic or planar anchorings on different types of interfaces\cite{li2018engineering,price2006anchoring,drawhorn1995anchoring}.
Moreover, in recent years, the photo-patterning technique has made it possible to design any arbitrary pattern of in-plane surface anchorings by using electromagnetic polarized light beams on a layer of photosensitive dyes\cite{guo2016higha,yu2019plasmonic,gibbons1991surface}.

Here we demonstrate the generation of different material flow regimes as controlled by rotating surface anchoring easy directions in nematic liquid crystal cells. We explore the role of co-rotating and counter-rotating anchoring at the two bounding surfaces, as well as the role of the relative phase. Different flow regimes are demonstrated, including no-net flow, oscillatory flow, steady flow, and pulsating flow.

\begin{figure}[t]
	\centering
	\includegraphics[scale=1]{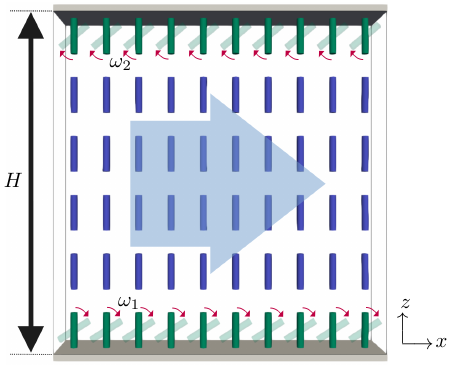}
	\caption{\textit{Generation of nematic flow by dynamic surface anchoring}. Surface anchoring easy axes orientate on the bottom and top surfaces with angular velocities $\omega_1$ and $\omega_2$, in $xz$ plane, resulting in a flow field, mostly in $x$ direction.} 
	\label{fig01}
\end{figure}

\section{Implementation of dynamic anchoring}

We used the tensorial description of the NLC to numerically simulate the flow inside the system, based on $\textbf{Q}$-tensor order parameter, with the largest eigenvalue and the corresponding eigenvector giving the local degree of order $S$ and the nematic easy axis $\hat{n}$, respectively.

Time evolution of the tensor order parameter and the flow field of the NLC are described using the Beris-Edwards approach \cite{BerisEdwards,Kleman.03},
\begin{eqnarray}
	&\dot{\textbf{Q}}=\Gamma \textbf{H} + \textbf{S}, \\
	& \rho\left[\partial_t v_i + \left(v_j\partial_j\right)v_i\right] = \partial_j \sigma_{ij}, \\
	& \partial_i v_i=0,
	\label{Beris}
\end{eqnarray}
where $\Gamma$ is called the rotational viscosity parameter.
The molecular field $\textbf{H}$ tensor leads the nematic director field to its equilibrium state where the Landau-de Gennes free energy \cite{deGennes.95} is minimum.
The generalized advection term $\textbf{S}$ couples the nematic orientational order with the material flow, $\rho$ is the fluid density, $\sigma_{ij}$ is the stress tensor as defined in Ref\cite{kos2020field}.
We use typical material parameters of characteristic nematics like 5CB: single elastic constant $L=4.8$ pN, rotational viscosity parameter $\Gamma=15 (\text{Pa.s})^{-1}$, 
Landau-de Gennes free energy parameters $A=-1.72 \times 10^{5}\text{J/m}^{3}$, $B=-2.12 \times 10^{6}\text{J/m}^{3}$, and $C=1.73 \times 10^{6}\text{J/m}^{3}$, which result in the nematic correlation length $\xi_\text{N}=2.3$nm.

We solve the Beris-Edwards equations using hybrid lattice Boltzmann algorithm \cite{PhysRevE.76.031921, Z.Kos.Nematodynamic}.
For that purpose, we consider a $100 \times 20 \times H$ grid box, with the the grid size $\Delta x = 1.5 \xi_\text{N}$, where the cell height $H$ varies between $50-200 \Delta x$. Periodic boundary conditions are considered in $x$ and $y$ directions.

As shown in Fig. \ref{fig01}, on the top and bottom cell surfaces, the surface anchoring easy axis $\hat{n}^0(t)=\left(\sin\theta,0,\cos\theta\right)$ is assumed to rotate in the $xz$ plane from planar to homeotropic to planar and so on, with constant angular velocities $\omega_{1}$ and $\omega_2$ on the bottom and top walls, respectively.
\begin{eqnarray}
	\theta_1 (t) = \omega_{1} t, \ \ \
	\theta_2 (t)= \omega_{2} t + \Delta\phi_A,
	\label{n1_n2}
\end{eqnarray}
where $\Delta\phi_A$ is the phase difference between the easy axis direction angles on the two surfaces.
We impose this easy axis boundary condition to the surface, as strong anchoring.
Specifically, we construct a time-variable surface anchoring imposed tensor $Q^{0}_{ij}(t)=S_{eq}(3\hat{n}^0_{i}(t)\hat{n}^0_{j}(t)-\delta_{ij})/2$, where $S_{eq}$ is the equilibrium NLC scalar order parameter, which we use as the boundary condition.
Here, we have to note that anchoring is the key factor in this mechanism as it is the source of the driving force, thus naturally, the effectiveness of flow driving reduces with reducing the anchoring strength.
This anchoring rotation generates a flow field in bulk, which is mostly in $x$ direction and the other components are negligible as they are 2-5 orders of magnitude smaller.
Throughout this article, the angular velocities $\omega_1$ and $\omega_2$ are given as a characteristic anchoring frequency $\omega_0 = 2 \pi /T_0$, where $T_0$ is $5\times 10^4$ simulation time steps defined as $\Delta t=0.025 \Delta x^2/(\Gamma L)$.
We start our simulations with a zero initial flow field and a uniform director field with small random deviations from the $z$ axis, which is perpendicular to the top and bottom solid cell walls.
The flow flux ($F(t)=\int \vec{v}\cdot\hat{x}dA$) is analyzed by Fourier series,
\begin{eqnarray}
	F(t) = a_0 + \sum_{n=1}^{\infty} a_n \cos(n \omega_1 t +\phi_n),
	\label{fourier}
\end{eqnarray}
where $a_0$ indicates the flux time average, $a_n$ and $\phi_n$ are the Fourier coefficients, and $\omega_1$ is the smaller angular velocity magnitude which in this work is always considered at the bottom surface.
An established way to characterize the relative strength between the viscous and elastic forces in liquid crystals is by the dimensionless Ericksen number $\text{Er}=\frac{vH}{\Gamma L}$\cite{BerisEdwards}.

\section{Results}

\begin{figure*}[!h]
	\centering
	\includegraphics[scale=0.85]{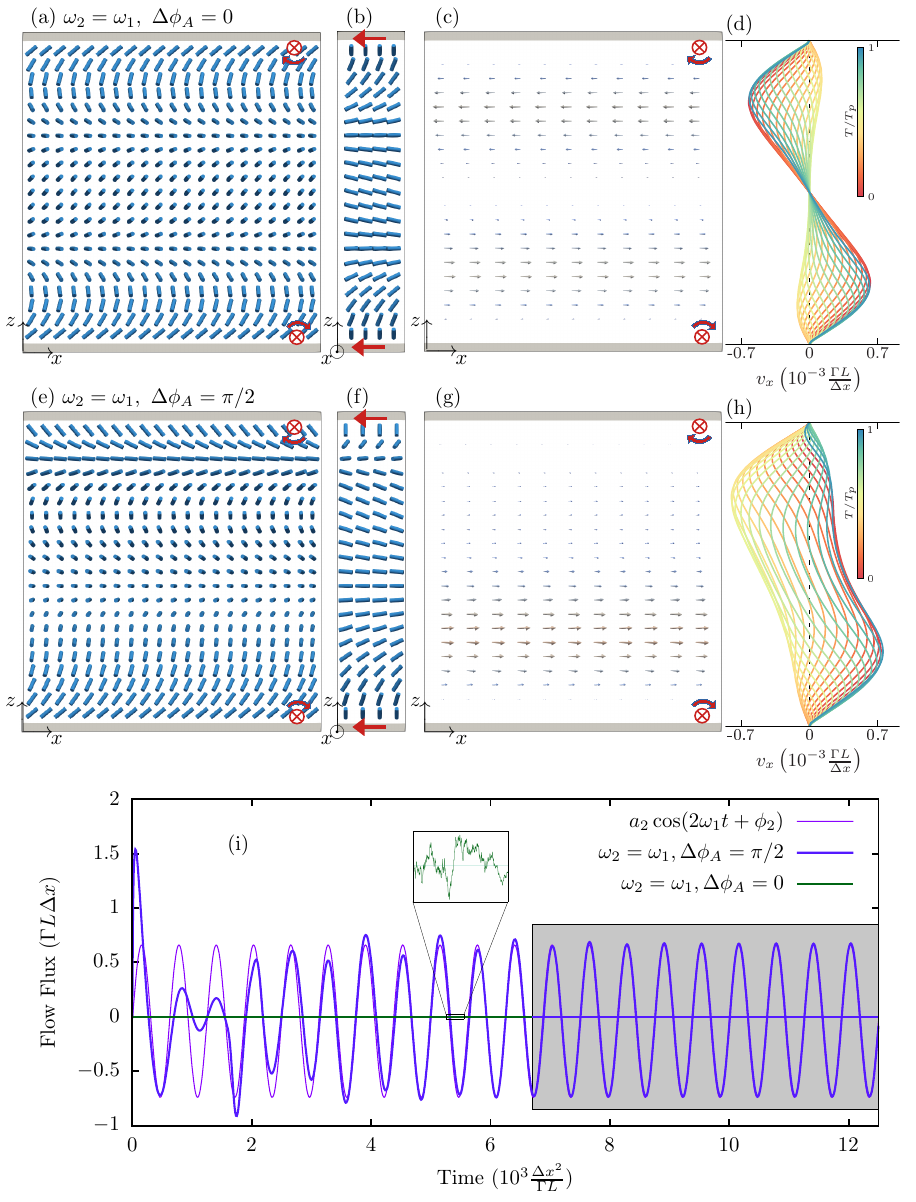}
	\caption{\textit{No net flow and oscillating flow regimes generated by co-rotating time-varying active surface anchorings.} Snapshots of the director and flow fields when $\omega_{2}=\omega_{1}=\omega_0$, with anchoring phase differences $\Delta\phi_A=0$ in (a)-(c) (first row), and $\Delta\phi_A=\pi/2$ in (e)-(g) (second row). Front and side views are displayed in the first and second columns. Anchoring angular velocity vectors $\omega_1$ and $\omega_2$ are shown with red arrows at the bottom and top surfaces. (c) and (g) show the local flow field at the same time as in (a)-(b) and (e)-(f), respectively. (d) and (h) show how the velocity field in the channel changes in one-period duration $T_p$;  maximum Ericksen number is $\text{Er} \approx 0.2$.  (i) Flow flux through a plane cross-section perpendicular to the $x$ axis in the middle of the system as a function of time. The gray box shows the steady state phase of the system, in which the time average of the flux is calculated. The non-zero Fourier coefficients are calculated as $a_2=0.70$ and $\phi_2=-0.52\pi$. The inset shows a close view of $\Delta\phi_A=0$ case where the fluctuations order of magnitude is $10^{-4}$. (See also supplementary videos 1 and 2).} 
	\label{fig02}
\end{figure*}

We explore the generation of material flow by rotating surface anchoring.
We start with co-rotating cases where the anchoring orientation rotates in the same direction and with the same angular velocities on both the top and bottom solid boundaries, $\omega_{2}=\omega_{1}$.
Figs. \ref{fig02} (a)-(c) and (e)-(g) show snapshots of the bulk director and flow velocity fields with two amounts of initial anchoring phase differences, $\Delta\phi_A=0,\pi/2$.
These snapshots are taken in the dynamic steady state, where the initial transient behaviors has died out and the system shows completely periodic behavior. The steady state is highlighted in Fig. \ref{fig02}(i).
Figs. \ref{fig02} (d) and (h) show $x$ component of the velocity field inside the bulk as a function of $z$, in one period duration time ($T_p=2\pi/\omega$). 

The director field mostly remains in the $xz$ plane near the top and bottom surfaces and rotates around the $y$ axis. By going farther into the bulk, the director field splays out of the plane, and then in the middle of the bulk, it shows a rotational motion on a cone around the $y$ axis with an easy angle $\psi$, which is a function of $z$ and has a minimum at $z=H/2$. 
The escape of the director field toward the $y$ axis helps the system avoid the growth of elastic distortions due to director winding. 

With $\Delta\phi_A=0$, the flow velocity field as shown in Fig. \ref{fig02}(a), is symmetric as well.
Here the flow is in the opposite directions in the top and bottom halves of the cell, where the flow is mostly in the left direction at the upper part and it is in the right direction in the lower part of the cell.
Therefore, the overall flow flux throughout any $yz$ cross-section of the cell ($\int \vec{v}\cdot\hat{x}dydz$) is always zero, as shown in Fig. \ref{fig02}(i).
Despite no (time-averaged) net flow, the local flow velocity vector oscillates in time, between its minimum and maximum.
Somewhat differently, for $\Delta\phi_A=\pi/2$, the vector field shows spatially asymmetric shape, as the flow vector field locally oscillates at upper and lower parts with a $\pi/2$ phase difference, which gives rise to an instantaneously non-zero oscillating back and forth flow flux with zero time average, as shown in Fig. \ref{fig02}(i).
The flow flux data follows an effective cosine function with angular velocity $2\omega_{1}$;
i.e.  note, twice as fast as the director orientation near the walls,
which is the result of nematic head-to-tail symmetry ($\hat{n}\equiv-\hat{n}$).
Namely, the nematic field close to the surfaces as the source of flow in the bulk returns to its initial orientation already after half a round rotation ($\theta \rightarrow \theta+\pi$) and drives the bulk fluid the same way as before (see supplementary video 1 and 2).
As a result, the flow velocity field has a period of half a round of the surface anchoring rotation.

\begin{figure*}[!h]
	\centering
	\includegraphics[scale=0.85]{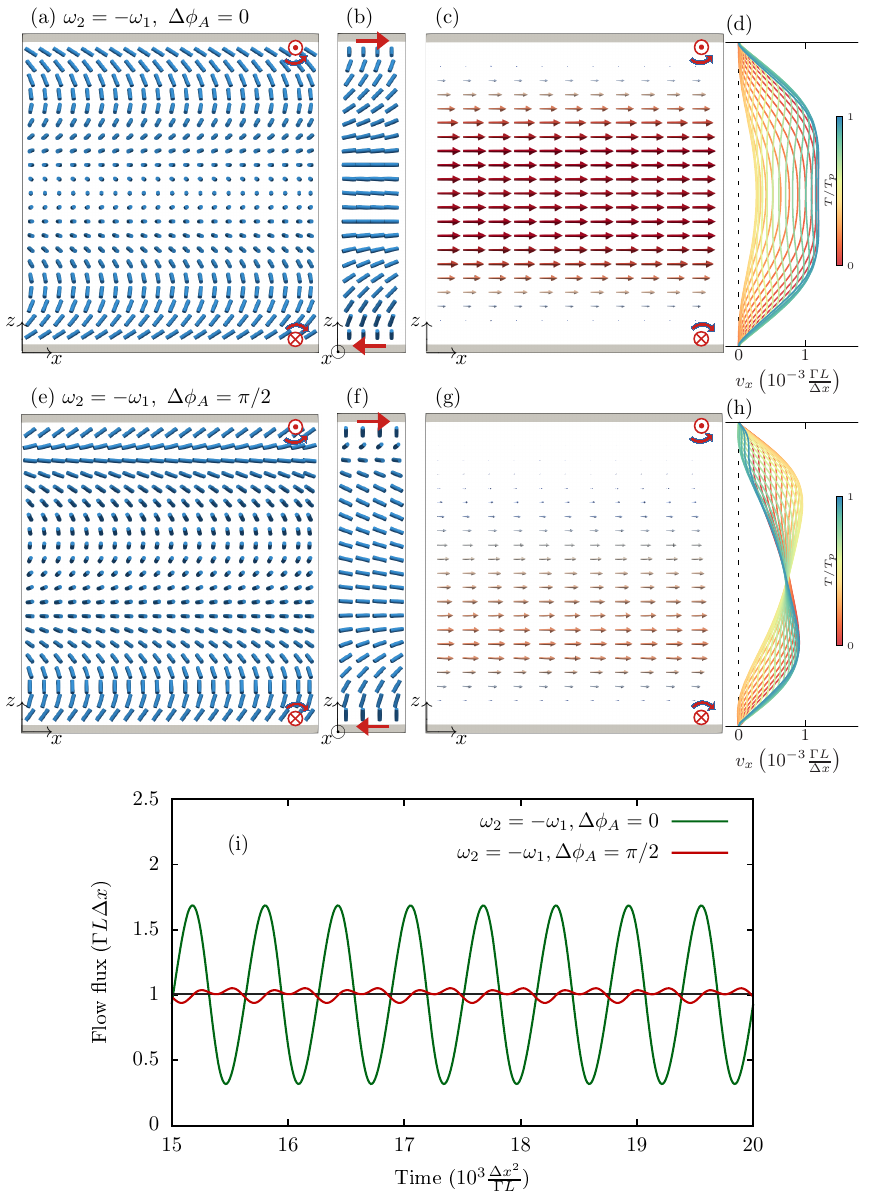}
	\caption{\textit{Steady flow regimes driven by counter-rotating time-varying active anchoring rotations.} Snapshots of the director and flow fields for $\omega_{2}=-\omega_{1}=-\omega_0$, with anchoring phase differences $\Delta\phi_A=0$ in (a)-(c) (first row), and $\Delta\phi_A=\pi/2$ in (e)-(g) (second row). Anchoring angular velocity vectors $\omega_1$ and $\omega_2$ are shown with red arrows at the bottom and top surfaces. (c) and (g) show the local flow field at the same time as in (a)-(b) and (e)-(f), respectively. (d) and (h) show how the velocity field in the channel changes in one-period duration $T_p$, where the maximum Ericksen number is $\text{Er} \approx 0.3$.  (i) Flow flux through a yz plane cross-section as a function of time. The non-zero Fourier coefficients calculated as $a_0=1.00$, $a_2=0.68$, $\phi_2=0.53\pi$, $a_4=0.04$ and $\phi_4=0.47\pi$ for $\Delta\phi_A=0$, and $a_0=1.00$, $a_2=0.03$, $\phi_2=-0.68\pi$, $a_4=0.03$ and $\phi_4=0.50\pi$ for $\Delta\phi_A=\pi/2$. (See also supplementary videos 3 and 4).} 
	\label{fig03}
\end{figure*}

\begin{figure*}[!h]
	\centering
	\includegraphics[scale=0.85]{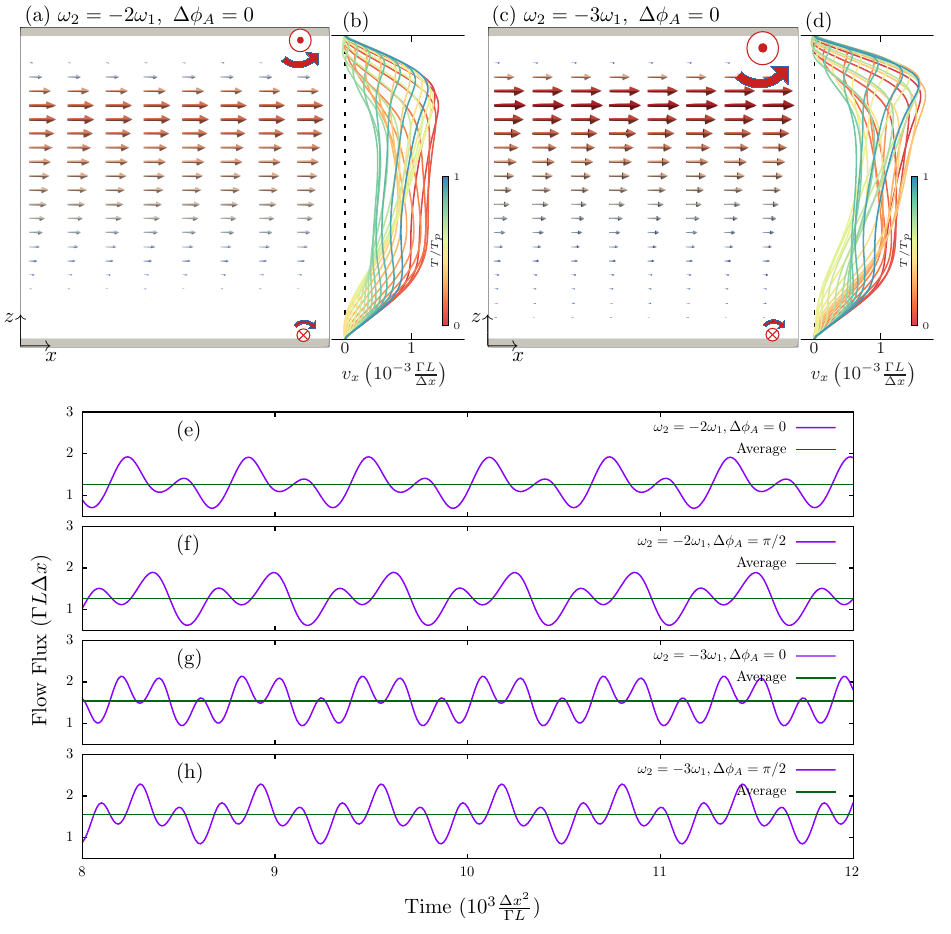}
	\caption{\textit{Tuning of steady flow profiles by different (counter-rotating) surface anchoring angular velocities.} Snapshots of the flow field for (a) $\omega_2=-2\omega_{1}$  and (c) $\omega_{2}=-3\omega_{1}$, and their corresponding variability in time within one period $T_p=\pi/\omega_0$ in (b) with $\text{Er} \approx 0.4$, and (d) with $\text{Er} \approx 0.45$. (e)-(h) Variability of the flow flux in time.  The non-zero Fourier coefficients are calculated as: $a_0=1.27, a_2=0.31, \phi_2=-0.52\pi, a_4=0.37, \phi_4=-0.64\pi$ for (e), $a_0=1.27, a_2=0.32, \phi_2=-0.52\pi, a_4=0.40, \phi_4=0.37\pi$ for (f), $a_0=1.54, a_2=0.33, \phi_2=-0.53\pi, a_6=0.39, \phi_6=-0.69\pi$ for (g), and $a_0=1.54, a_2=0.33, \phi_2=-0.52\pi, a_6=0.39, \phi_6=0.31\pi$ for (h) .} 
	\label{fig04}
\end{figure*}

In order to observe the net flux, we choose different anchoring angular velocities on the top and bottom surfaces.
In Fig. \ref{fig03}, we show an example where the director on the two boundaries rotate with the same angular velocity but in opposite directions, $\omega_{2}=-\omega_{1}$.
The director field behavior is mostly similar to the symmetric case of in Fig. \ref{fig02}, except that it is mirror-symmetric as the bending directions are the same at the upper and lower halves of the cell.
By using $\Delta\phi_A=0$, and counter-rotating anchoring angular velocities, as shown in Fig. \ref{fig03}(a)-(d), effectively, the same synchronized elastic distortions emerge and importantly, with the same direction of the backflow generated material flow driving. The local flow field oscillates all over the bulk, and the magnitude is always largest in the middle, which results a sinusoidal profile of the flow flux which is shown in Fig. \ref{fig03}(i).
For $\Delta\phi_A=0$, the flow flux has the form of the second term of Fourier series with the angular velocity $2\omega_{1}$ in accordance with Fig. \ref{fig03}(i).
For $\Delta\phi_A \neq 0$ the flux time profile becomes more complicated as the fourth fourier terms appear.
The phase difference does not affect the frequency of the flow flux or its time average. But it does decrease the magnitude of the flow beats. Consequently, the phase difference between the time-variable surface anchorings can be used as a control parameter for tuning the flow smoothness and the overall profile.

\begin{figure*}[!h]
	\centering
	\includegraphics[scale=0.85]{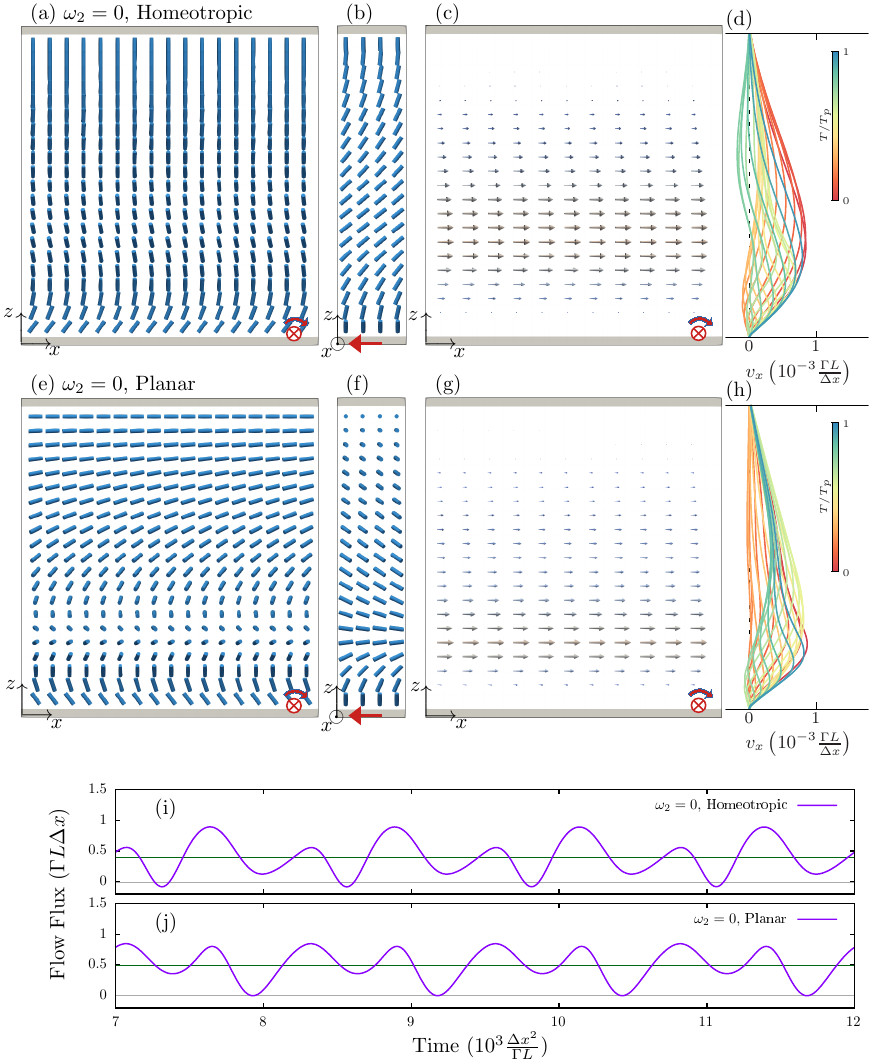}
	\caption{\textbf{Pulsating flow regimes driven by single-surface time-variable active anchoring.} Snapshots of the director and flow field when the anchoring is fixed on the top surface $\omega_{2}=0$ and $\omega_1=\omega_0$ on the bottom. For the static top surface, homeotropic (a)-(d) and non-degenerate planar (e)-(h) is assumed; corresponding flow profiles are shown in (d) and (h) respectively as functions of time, where the maximum Ericksen number is $\text{Er} \approx 0.3$. Flow flux as a function of time when homeotropic (i) and planar (j) anchorings on the static surface with Fourier coefficients $a_0=0.39, a_1=0.18, \phi_1=-0.28\pi, a_2=0.33, \phi_2=-0.48\pi, a_3=0.07, \phi_3=-0.20\pi, a_4=0.03, \phi_4=0.28\pi$ for (i), and $a_0=0.50, a_1=0.18, \phi_1=0.40\pi, a_2=0.30, \phi_2=-0.50\pi, a_3=0.06, \phi_3=-0.82\pi, a_4=0.02, \phi_4=0.70\pi$ for (j). (See also supplementary videos 5 and 6).} 
	\label{fig05}
\end{figure*}

Figure \ref{fig04} shows the adjustment of the flow profiles and the flow magnitude taking different (counter-rotating) angular velocities of the surface anchoring easy axes on the two surfaces.
Figs \ref{fig04}(a)-(d) show that the flow velocity oscillates with a larger magnitude near the top surface, where the anchoring rotation is faster, which notably, also allows the flow to steer to the top or bottom of the channel.
Moreover, by increasing $\omega_2$ the flow amplitude and also the flux average increases, as shown in Figs. \ref{fig04}(e)-(h). 
The overall period of the system is still determined by the slower surface anchoring on the bottom, whereas the faster rotation at the top leads to the flux time behavior as higher order terms in the Fourier series, with exact values of the coefficients determined by the anchoring phase difference $\Delta\phi_A$.

In Fig. \ref{fig05} we have studied the case of static surface anchoring on the top surface, $\omega_2=0$, while we kept the dynamic anchoring on the bottom, $\omega_1=\omega_0$. Specifically, the standard homeotropic and planar anchorings are considered at the static top surface.
Figs. \ref{fig05}(d), (h), (I), and (j) show that this static-dynamic time-variable anchoring creates a pulsating flow, where the flux periodically stops or even draws a little bit back at specific moments and then drives on again.
The next significant difference here appears in the leading frequency of the flow (and flow flux), which is found to be the same as the anchoring frequency $\omega_1$ (and not $2\omega_1$ as before with two dynamic surfaces).
The fixed anchoring in both homeotropic and planar conditions does not allow the system to recover its initial state when the anchoring direction rotates by only half a turn (i.e. by $\pi$). See supplementary videos 5 and 6.
Therefore, after a $\pi$ rotation of the director on the bottom surface, the director field undergoes different configurations, resulting in different backflow coupling and thus different material flow.
We note that the $2\omega_1$ frequency is still present in the Fourier series of the flow flux as the second-order term.
The time average of the flux is expectedly different in homeotropic and planar anchorings due to the different bulk distortions and different backflow effects.

\begin{figure*}
	\centering
	\includegraphics[scale=0.8]{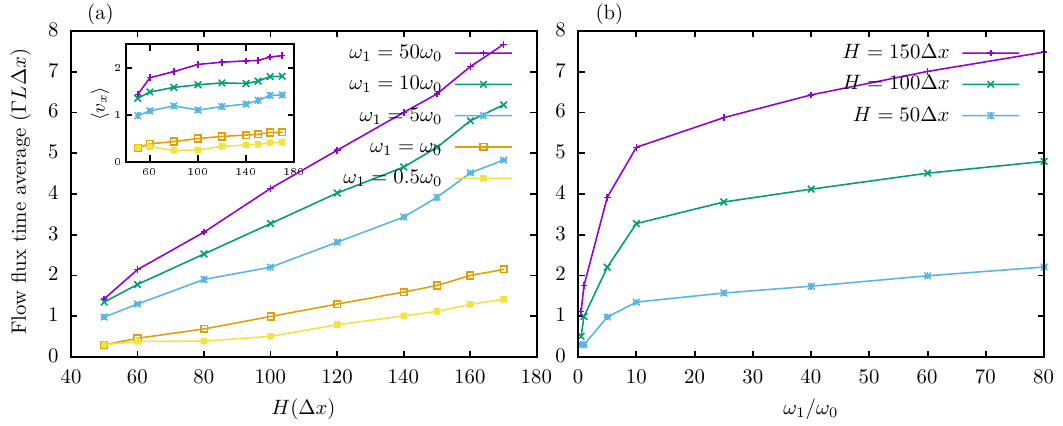}
	\caption{Time average of the flow flux as a function of the cell height $H$ (a), and anchoring frequency $\omega_{1}$ (b) for counter-rotating driving with $\omega_2=-\omega_1$. The inset in (a) shows the flow flux density, defined as the flow flux time average divided by the cross-section area ($a_0 / A$) in units of $10^{-3} \frac{\Gamma L}{\Delta x} \approx 20 \mu\text{m/s}$.} 
	\label{fig06}
\end{figure*}

Figs. \ref{fig06}(a) and (b) show the flow flux as dependent on the cell thickness $H$ and the magnitude of anchoring angular velocity when $\omega_2=-\omega_1$.
The net flux increases with the anchoring angular speed for small $\omega_1$.
For large $\omega_1$, the backflow effect becomes less efficient since the high speed of the anchoring overcomes the relaxation time of the nematic fluid, and as a result, the net flux approaches a saturated level. The net flux also increases by the cell thickness $H$, as shown in Fig. \ref{fig06}(a). The inset in \ref{fig06}(a) shows that the cell thickness has some effect on the average flow velocity across the channel bulk, calculated as $<v_x>=a_0/A$, where $a_0$ is the zero-order Fourier flow flux coefficient or the time average of the flow flux and $A$ is defined as the area of the cell cross-section.
Here, note that although the flow flux magnitude increases with the cell thickness, its qualitative behavior does not change much, and even the average flow velocity in the inset of Fig. \ref{fig06}(a) varies only weakly with the width of the channel. This can be a result of the fact that the considered nematic orientational profiles are effectively scalable as only conditioned by the director's elasticity and not nematic scalar order.

\section{Discussion and Conclusion}
We can propose two possible ways to bring tunable dynamic surface anchorings into operation in an experiment:
(i) Multi-step photo-alignment; Surface anchoring can be controlled and switched by photoresponsive carbohydrate-based surfactants\cite{PhotoresponsiveCarbohydrate,noh2018sub} or azobenzen-based elastomers grafted onto the surface\cite{wang2016self,xue2015light,quint2015all,seki2013versatility}.
Such elastomers switch their bending under the ultraviolet wave and back by visible light, which results in surface anchoring switching because of their interactions with NLC molecules near the surface.
By switching the surface anchoring over a multistep cycle, the dynamic anchoring can be -at least in principle- available.
Such photoisomerizations occur on a time scale of picoseconds\cite{C6CP08461C}, which is quite fast, but the reorientation of the NLC molecules could seem to be slow due to the high orientational viscosity, since the time scale of collective nematic response to elastic distortions is $\tau_\text{dir}=\frac{H^2}{\Gamma L}\sim 1\text{s}$ for a typical experimental length scale of $H=10\mu \text{m}$.
But as a limited dynamic region close to the surface is sufficient for flow generation -thanks to the escaping of the bulk director field in $y$ direction- a more relevant time scale here would be multiples of $\tau_\text{N}=\frac{\xi^2_\text{N}}{\Gamma L} \approx 70 n\text{s}$, which illustrates the nematic elastic response time within the correlation length.
(ii) Electro-optical meta-surfaces\cite{kossyrev2005electric,10.1063/1.2837099,minovich2012liquid,decker2013electro,buchnev2013electro,wu2020liquid} are 2D nano-meter size grids of electrodes, with the ability to control tangential fields locally on the surface, which coupling with nematic molecular orientations, make them very agile and rapid tools for controlling and programming the dynamic surface anchoring on the boundaries of NLCs, with reorientation time scale of milliseconds.
Again, with proper design and engineering, time-variable easy axes modulation could potentially be realized.

In conclusion, we demonstrate the generation of material flow in flat cell geometry of nematic complex fluids as driven by time-variable active surface anchoring. 
Our results show that the co-rotating or counter-rotating anchorings lead to different flow regimes.
If we look at the results shown in Figs. \ref{fig02}-\ref{fig05}, effectively, we notice that the local flow direction is always roughly determined by the angular velocity vector of the nearby surface anchoring through the right-hand law. 
As a result, for co-rotating anchorings where the angular velocity has the same magnitude and direction, the top and bottom surfaces effectively cancel out each other's effect and result in a zero time-averaged net flux, while non-zero oscillating flux (with zero time-average) can emerge by imposing a phase shift between the angles of rotation of the two surfaces. 
Differently, for counter-rotating time-variable active surface anchoring, when the anchorings on the top and bottom rotate with the same angular velocity but in opposite directions, effectively, both active surfaces work up together which leads to a directional net material flow.
Moreover, by choosing different rotation speeds at the top and bottom, the flow profile can be steered towards the top or bottom surface as a larger flow is realized near the wall with faster anchoring rotation, which can be a control parameter for flow steering.
In parallel, we show that the phase shift $\Delta \phi_A$ can be used to tune the flow profiles as well as the exact flow flux time dependence. 
Fourier analysis of the flow flux of co-rotating and counter-rotating surface anchorings shows that only even Fourier terms contribute to the flow flux, and as a result, the overall frequency of the flow flux is twice as large as the smallest anchoring frequency.
This constraint changes when one of the surfaces is fixed and the other is rotating, where both odd and even terms emerge in the flow flux.
Pulsating flow flux is a characteristic behavior of this static-dynamic time-variable surface anchoring regime, where the dynamic surface anchoring is applied only on one side.
In our simulations, the Ericksen number varies between less than 0.1 for thin layers with small rotation angular velocities and 0.6 for thicker layers with faster anchoring rotations. This number is comparable to a typical maximum velocity of $v=6\mu\text{m/s}$ in experiment\cite{PhysRevLett.99.127802} and a typical cell thickness of $H=10\mu\text{m}$, with the same rotational viscosity and elastic constant used in this article (see section 2).
We do not see any topological defects in our simulations in the steady state as shown in Figs. \ref{fig02}-\ref{fig05}. We believe there are two reasons for the absence of the defects here: (i) simple 2D-infinite plane geometry for the boundaries and (ii) uniform anchoring rotation with no dependency on position. Of course, by choosing more complicated geometries or scenarios for the dynamic surface anchoring driving, the emergence of topological defects and consequently, more complicated flow fields is expected.

In this work, we focused on the ability and possible modalities of the surfaces to drive and control the microfluidic flow of complex nematic fluids which would broaden the options of pumping NLC flow in the experiment as a complement to other propositions for driving NLC flow in the bulk for example by using electric or optical fields\cite{kos2020field}.
For example, the distinct difference of this work is that the flow driving does not depend on any material dielectric properties (e.g. refractive indices, dielectric constant, etc.) and could be applied also to zero-birefringence nematic fluids.
Secondly, the concept of flow driving through time-variable surface anchoring could -by using different surface anchoring driving mechanisms- be transferable to other nematic material length scales such as colloidal liquid crystals which could be much more challenging with optical (size) or electric field effect (homogeneity of the field).
Finally, the effectively surface-imposed dynamic control could also be transferred to different -for example, active- materials (i.e. beyond molecular liquid crystals) where bulk control of the materials is currently much less established or available.
More generally, this work contributes to the development of novel material driving routes in topological soft matter and synthetic active materials.

\section*{Author Contributions}
Both authors contributed equally to this work.

\section*{Conflicts of interest}
There are no conflicts to declare.

\section*{Acknowledgements}
Authors acknowledge helpful discussions with \v{Z}iga Kos. 
The authors acknowledge funding from the Slovenian research agency ARIS grants P1-0099, N1-0195, J1-50004, J1-2462, and J1-50004. This result is part of a project that has received funding from the European Research Council (ERC) under the European Union’s Horizon 2020 Research and Innovation Program (Grant Agreement No. 884928-LOGOS).



\balance


\bibliography{Active_Anchoring} 
\bibliographystyle{rsc} 

\end{document}